# Serial and parallel combinations of diodes: equivalence formulae and their domain of validity

Raymond Laagel and Olivier Haeberlé
Université de Haute-Alsace, Institut Universitaire de Technologie de Mulhouse
Groupe Lab.El - Equipe EEA, 61 rue Albert Camus 68093 Mulhouse Cedex FRANCE
olivier.haeberle@uha.fr

**ABSTRACT :** Serial and parallel combinations of diodes are studied by first establishing approached equivalence formulae. Sorting diodes enlights the true dispersion of their characteristics. This simple exercice helps students to realize that "identical" components may have in fact slightly different behaviour, especially in analogic electronic. Using very common components like diodes and simple electronics assembly, students can then face and test the notion of practical domain of validity for a theoretically established formula. Using diodes is especially interesting to introduce these notions, because the domain of application of the proposed formulae strongly depends on the components combination (parallel or serial), because of their non-linear behaviour.

## 1 INTRODUCTION

When using diodes in parallel or serial combination, one should take care to choose diodes with similar characteristics in order to balance the current in the diodes (parallel combination) or the voltage (series combination). To get diodes with similar characteristics out of a manufacturer's set, one has to sort the diodes according to some criteria. Then a formula giving the diode equivalent to a combination of diodes can be used.

We propose to familiarise the student with the idea of validity of approximate expressions by studying a very simple and common non-linear component. By first sorting some diodes through a simple electronic assembly, they realize that manufacturers do have fabrication tolerances. Then using a model of the diode, the equivalence formulae for the serial and parallel combinations are given. The domain of validity of the established formulae is studied. Using the selected diodes, it is easy to verify the obtained results.

## 2 DIODE MODEL

Several methods are available to sort a set of diodes. For example, one can measure the voltage for a fixed diode current. Another and more accurate method is to determine the parameters of a model of the diode. The simplest is the exponential model giving the diode current:

$$I = I_S \left[ \exp\left(\frac{eV}{kT}\right) - 1 \right] \quad (1)$$

where $I_S$ is the saturation current, V the voltage applied to the diode (anode to cathode), k the Boltzmann constant and T the temperature. This method is more complicated than the first one, but has the advantage that the model is valid over several decades of variation for the diode current. A justification of the simple diode equation Eq. (1) has been given in Ref. [1]. We propose to show students how to establish equivalence formulae in the case of parallel and serial combinations of two diodes obeying this simple equation.

Working at fixed room temperature (which is a realistic hypothesis), one sets $V_T = \frac{kT}{e}$. As a consequence, Eq. (1) becomes (Refs. [2,3]):

$$I = I_S \left[ \exp\left(\frac{V}{V_T}\right) - 1 \right] \quad (2)$$

With this simple model, the diode is modeled with two parameters only:
- $V_T$, which has the dimension of a voltage. The experimental value of $V_T$ varies with the type of semiconductor constituting the junction (Refs. [4,5]).
- $I_S$, the theoretical saturation current of the diode.

Throughout this work, we focus our attention on silicon forward biased diodes, and the above expression Eq. (2) simplifies if $\exp\left(\frac{V}{V_T}\right) \gg 1$ to obtain:

$$I \approx I_S \exp\left(\frac{V}{V_T}\right). \quad (3)$$

From Eq. (2), one can also express the voltage as:

$$V = V_T \left[ \ln\left(\frac{I}{I_S}\right) + 1 \right]. \quad (4)$$

Similarly to Eq. (2), Eq. (4) simplifies if $\ln\left(\frac{I}{I_S}\right) \gg 1$ and one has:

$$V \approx V_T \ln\left(\frac{I}{I_S}\right).$$

(5)

At T = 20°C (293 K) and for silicon diodes, the relative error $\frac{\Delta I}{I}$ one makes when using Eq. (3) instead of Eq. (2) is in practice lower than 1 ‰ as soon as V ≥ 350 mV. Similarly, the relative error $\frac{\Delta V}{V}$ one makes when using Eq. (5) instead of Eq. (4) is lower than 1 ‰ when $I \geq 190\ I_S$. Therefore Eqs. (3) and (5) will be used in the following.

If one could get strictly identical diodes, the equivalent models for these two diodes would be very easy to establish. For two serial mounted diodes, one would have:

$$V^* = 2\ V_T\ \ln\left(\frac{I}{I_S}\right)$$

(6)

and for two diodes in parallel mounting:

$$I^* = 2\ I_S\ \exp\left(\frac{V}{V_T}\right)$$

(7)

In fact, as it is never possible to find strictly identical diodes, from a certain set of diodes, one should choose those with the closest $I_S$ and $V_T$ parameters. The experimental determination of these parameters is done using the static characteristic of the diode. This is a simple task, not out of the ordinary for (first year) students doing electronic lab courses (Ref. [6]).

Figure 1 illustrates the dispersion of $I_S$ and $V_T$ (when plotting $I_S = f(V_T)$) one can get out of a set of standard diodes (1N4002). It is worth noting that the dispersion is actually much larger for $I_S$ than for $V_T$. It is clear than a large majority of the diodes belong to a domain defined by 40 mV ≤ $V_T$ ≤ 50 mV and 2 nA ≤ $I_S$ ≤ 6 nA.

The relative variations of $V_T$ and $I_S$ can be written as:

$$x = \frac{\Delta V_T}{V_T} \text{ and } y = \frac{\Delta I_S}{I_S}.$$

With careful selection it is possible to find diodes whose characteristics not being equal are nevertheless very similar $x \approx \pm 2.10^{-3}$, $y \approx \pm 3.10^{-2}$. We propose to study the limits for which it is possible to establish a simple expression valid for each type of diode combination and of more general validity than Eqs. (6)-(7). Figure 2 details the notation we adopt.

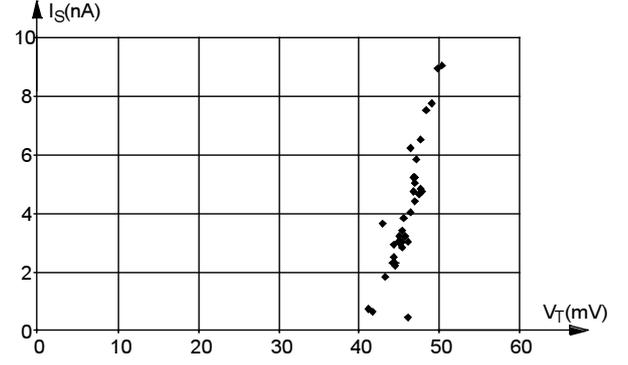

*fig 1 : Plot of $I_S = f(V_T)$ for a set of 1N4002 diodes. Note that the dispersion is actually much larger for $I_S$ than for $V_T$ but that most diodes are characterized by 40 mV ≤ $V_T$ ≤ 50 mV and 2 nA ≤ $I_S$ ≤ 6 nA.*

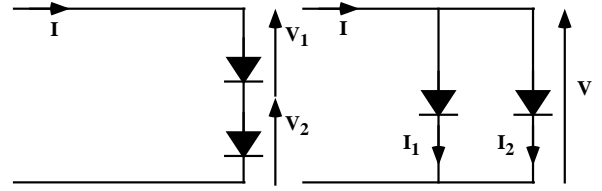

*fig 2 : Notations for the serial (a) and parallel (b) combinations of diodes.*

## 3 SERIAL COMBINATION

We define $V_1 = V_{T1} \ln\left(\frac{I}{I_{S1}}\right)$ and $V_2 = V_{T2} \ln\left(\frac{I}{I_{S2}}\right)$ as the voltages of the two diodes (see Fig. 2(a)). In the case of serial combination, as the same current I flows through both diodes, the voltage for the set is obviously:

$$V = V_1 + V_2 = V_{T1} \ln\left(\frac{I}{I_{S1}}\right) + V_{T2} \ln\left(\frac{I}{I_{S2}}\right)$$

(8)

One sets $V_{T0} = \sqrt{V_{T1} V_{T2}}$ and $I_{S0} = \sqrt{I_{S1} I_{S2}}$. From Fig. 1, we will take $V_{T0}$ = 45 mV and $I_{S0}$ = 4 nA (mean values of $V_T$ and $I_S$ for our set of diodes). Then Eq. (8) can be rewritten as:

$$V = V_1 + V_2 = V_{T0}\left(\sqrt{\frac{V_{T1}}{V_{T2}}}\right)\ln\left(\frac{I}{I_{S0}}\sqrt{\frac{I_{S2}}{I_{S1}}}\right)$$

$$+ V_{T0}\left(\sqrt{\frac{V_{T2}}{V_{T1}}}\right)\ln\left(\frac{I}{I_{S0}}\sqrt{\frac{I_{S1}}{I_{S2}}}\right) \quad (9)$$

or:

$$V = V_{T0}\left[\left(\sqrt{\frac{V_{T1}}{V_{T2}}} + \sqrt{\frac{V_{T2}}{V_{T1}}}\right)\ln\left(\frac{I}{I_{S0}}\right)\right.$$

$$\left. - \frac{1}{2}\ln\left(\frac{I_{S1}}{I_{S2}}\right)\left(\sqrt{\frac{V_{T1}}{V_{T2}}} - \sqrt{\frac{V_{T2}}{V_{T1}}}\right)\right] \quad (10)$$

Obviously, this formula does not change if one swaps both indices 1 and 2, namely one swaps both diodes. In order to take into account the dispersion of the diode characteristics, one can rewrite Eq. (10) as a function of the relative errors on the voltage and current, x and y, respectively:

$$V = V_{T0}\left[\left(\sqrt{1+x} + \frac{1}{\sqrt{1+x}}\right)\ln\left(\frac{I}{I_{S0}}\right) - \frac{1}{2}\left(\sqrt{1+x} - \frac{1}{\sqrt{1+x}}\right)\ln(1+y)\right] \quad (11)$$

Equation (11) reflects the fact that V is a function of the current I passing through the diodes, but also of the three functions of x and y:

$$A_1(x) = \sqrt{1+x} + \frac{1}{\sqrt{1+x}}$$
$$A_2(x) = \sqrt{1+x} - \frac{1}{\sqrt{1+x}}$$
(12)  $$B(y) = \ln(1+y)$$

so that:

$$V = V_{T0}\left[A_1(x)\ln\left(\frac{I}{I_{S0}}\right) - \frac{1}{2}A_2(x)B(y)\right] \quad (13)$$

For small relative variations x and y, one gets the approximate expression:

$$V_a = 2\ V_{T0}\ \ln\left(\frac{I}{I_{S0}}\right)$$

(14)
The relative error between the exact expression Eq. (13) and the approximation Eq.(14) is given by:

$$REs = \frac{V - V_a}{V} = 1 - \frac{2\ln\left(\frac{I}{I_{S0}}\right)}{A_1(x)\left[\ln\left(\frac{I}{I_{S0}}\right) - \frac{A_2(x)B(y)}{2A_1(x)}\right]} \quad (15)$$

Taking into account the earlier remarks, it appears that the larger the current I (compared to $I_{S0}$), the smaller the error REs. Furthermore, we have seen that $I > 190\ I_S$ is necessary for Eq. (5) to be valid within 1 ‰. This condition is satisfied for a current as low as I = 2μA with a saturation current of the order of 10 nA or less for the 1N4002 diodes.

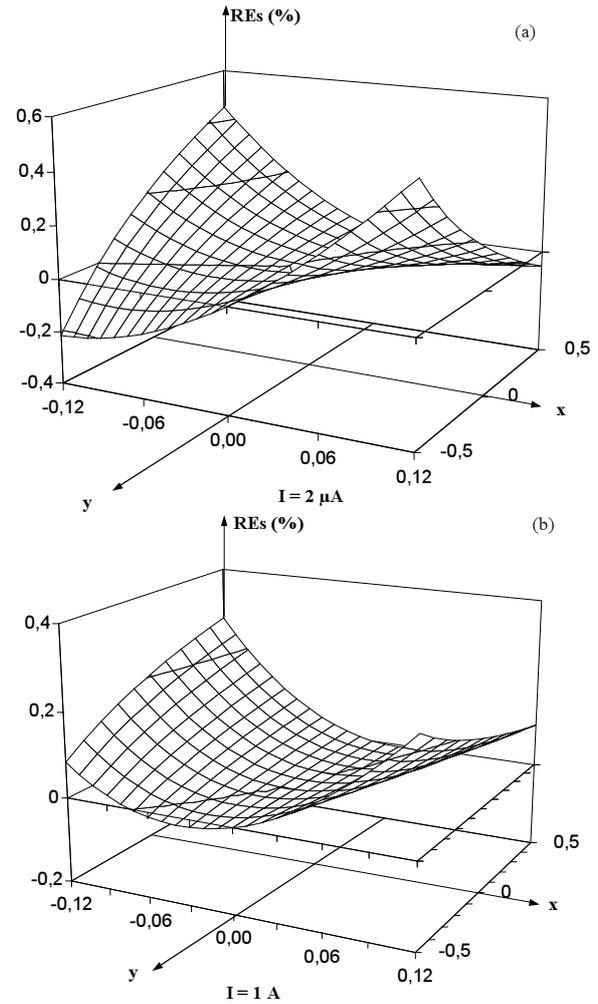

fig 3 : Relative error REs for the voltage in the case of two diodes in serial combination. (a): I = 2 μA; (b): I = 1 A. Note the low errors REs (≤ 0.6 %) even for large variations of x (± 12 %) and y (± 50 %) for this type of combination.

Figure 3 presents the relative error REs as a function of x and y for two values of the current I = 2 μA (lower limit of conduction) and I = 1 A (higher limit of use for the 1N4002). It is worth noting that even for a variation of almost 6 decades of the current I, the relative error REs is always lower than 0.6 % even for variations of $V_T$ and $I_S$ as large as 12 % and 50 %, respectively. As a consequence, the approximation given by Eq. (14) can be considered as good in a broad range of use.

**4 PARALLEL COMBINATION**
The theoretical developments are very similar to the case of the serial combination. The currents in the diodes, as a function of the common voltage are defined by $I_1 = I_{S1}\exp\frac{V}{V_{T1}}$ and $I_2 = I_{S2}\exp\frac{V}{V_{T2}}$ (see Fig 2(b)). One has then:

$$I = I_1 + I_2 = I_{S1} \exp \frac{V}{V_{T1}} + I_{S2} \exp \frac{V}{V_{T2}}$$

(16)

Inserting the term $I_{S0} = \sqrt{I_{S1} I_{S2}}$, one gets:

$$I = I_{S0} \left[ \left( \sqrt{\frac{I_{S1}}{I_{S2}}} \right) \exp \frac{V}{V_{T1}} + \left( \sqrt{\frac{I_{S2}}{I_{S1}}} \right) \exp \frac{V}{V_{T2}} \right] \quad (17)$$

Setting $\frac{2}{V_S} = \frac{1}{V_{T1}} + \frac{1}{V_{T2}}$ and $\frac{1}{V_d} = \frac{1}{V_{T1}} - \frac{1}{V_{T2}}$ one obtains finally:

$$I = I_{S0} \left[ \left( \sqrt{\frac{I_{S1}}{I_{S2}}} \exp V \left( \frac{1}{V_S} + \frac{1}{2 V_d} \right) \right) + \left( \sqrt{\frac{I_{S2}}{I_{S1}}} \exp V \left( \frac{1}{V_S} - \frac{1}{2 V_d} \right) \right) \right] \quad (18)$$

and one recognizes:

$$I = 2 I_{S0} \exp \left( \frac{V}{V_S} \right) \cosh \left[ \frac{1}{2} \left( \frac{V}{V_d} + \ln \frac{I_{S1}}{I_{S2}} \right) \right]$$

(19)

Again, this expression does not change if one swaps both indices 1 and 2, namely one swaps both diodes. As for the case of the serial combination, we will set an approximate expression for Eq. (19) and determine its domain of application. With this aim, we express $V_S$ and $V_d$ as a function of $V_{T0}$, x and y:

$$V_S = \frac{2 V_{T1} V_{T2}}{V_{T1} + V_{T2}}$$

or $V_S = 2 \frac{V_{T0}}{\sqrt{1+x} + \frac{1}{\sqrt{1+x}}} = 2 \frac{V_{T0}}{A_1(x)}$

(20)

Similarly, we have:

$$V_d = \frac{V_{T1} V_{T2}}{V_{T2} - V_{T1}} = \frac{-V_{T0}}{\sqrt{1+x} - \frac{1}{\sqrt{1+x}}} = -\frac{V_{T0}}{A_2(x)}$$

(21)

so that:

$$I = 2 I_{S0} \exp\left( \frac{V}{V_{T0}} \frac{A_1(x)}{2} \right)$$
$$\times \cosh\left[ \frac{1}{2} \left( -\frac{V}{V_{T0}} A_2(x) + B(y) \right) \right]$$

(22)

For small relative variations x and y, one obtains the approximate expression:

$$I_a = 2 I_{S0} \exp \left( \frac{V}{V_{T0}} \right)$$

(23)

The relative error between the exact expression Eq. (22) and the approximation Eq. (23) is given by:

$$REp = \frac{I - I_a}{I} = 1 - \frac{1}{\exp\left( \frac{V}{V_{T0}} \left( \frac{A_1(x)}{2} - 1 \right) \right) \cosh\left[ \frac{1}{2} \left( -\frac{V}{V_{T0}} A_2(x) + B(y) \right) \right]}$$

(24)

The lower the voltage V (compared to $V_{T0}$), the lower the error REp. Figure 4 shows the relative error REp for the voltage as a function of x and y for V = 0.35 Volt (lower limit of conduction) and V = 1 Volt (higher limit of use).

Note that in both cases, the error REp grows rapidly to become superior to 10% and for the same relative variations x and y, the error grows faster when V increases. As a consequence, Eq. (23) rapidly becomes an unsuitable approximation for the total current I.

It is then interesting to determine the domain determined by the relative variations of the voltage x and current y for which REp is lower than a fixed value p (in percentage). This domain is a function the voltage V applied to the diodes. We then define:

$$f(V/V_{T0}, x, y) = \exp\left( \frac{V}{V_{T0}} \left( \frac{A_1(x)}{2} - 1 \right) \right) \cosh\left( \frac{-V}{2 V_{T0}} A_2(x) - B(y) \right)$$

(25)

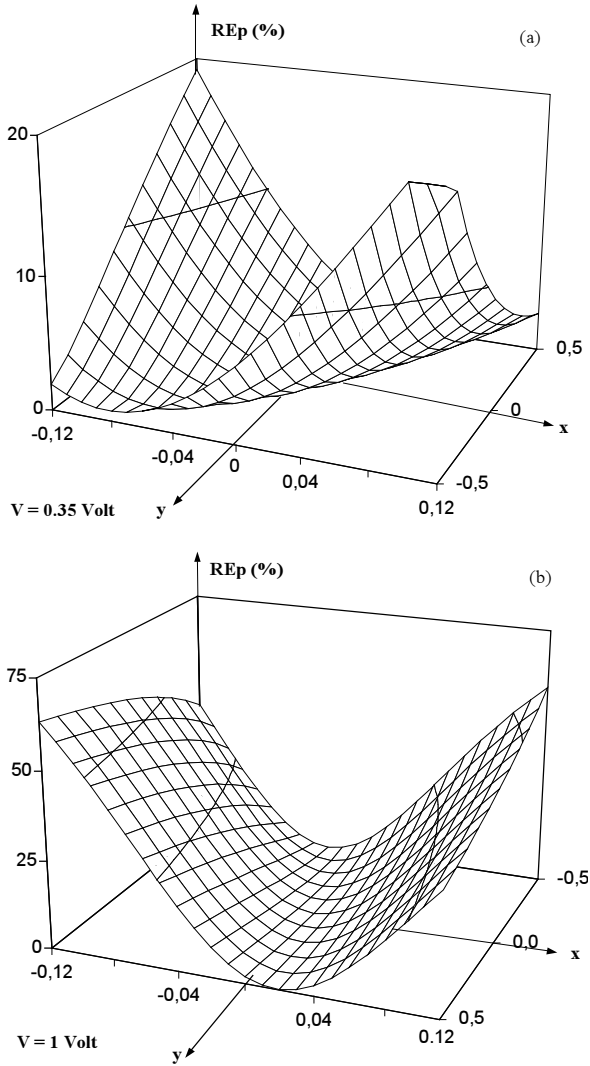

fig 4 : Relative error REp for the voltage in the case of two diodes in parallel combination. (a): $V = 0.35$ V, (b): $V = 1$ V. The error REp grows very fast to be larger than 10 %: $I_a$ is very rapidly not a good approximation for the total current.

so that $REp = 1 - \dfrac{1}{f(V/V_{T0},x,y)} = \dfrac{p}{100}$.

One has to solve the latter equation to determine the couples (x,y) which limit the domain of validity of $I_a$ as given by Eq. (24) for a relative error of p %.
As an illustration we choose a value of p = 1 %, and for several values of V, we determine the zone of validity of $I_a$.

Figure 5 shows the results obtained for V = 0.35 Volt, V = 0.68 Volt and V = 1 Volt. It is interesting to note that even for the most used case V = 0.68 Volt, the domain of validity is very narrow, compared to the serial combination for which large variations of x and y are possible.

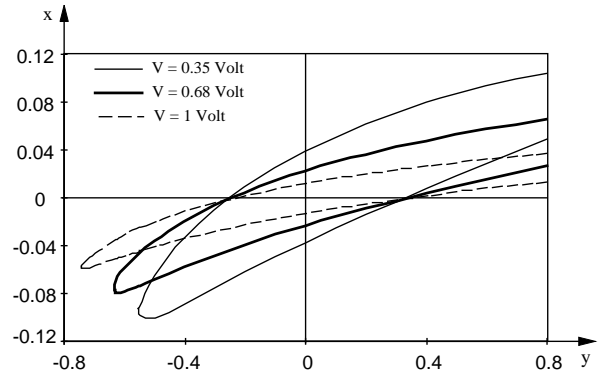

fig 5 : Domain of validity of the expression giving the approximate currrent $I_a$ (Eq. (24)) for a relative error of 1 % at V = 0.35 Volt, V = 0.68 Volt (the most common case of use) and V = 1 Volt..

## 5 CONCLUSION
We have shown that the approximate expression giving the voltage for two diodes in serial combination is of much wider validity than the corresponding approximate expression in the case of a parallel combination. This is easily understandable if one notices that in the serial combination, one adds two logarithmic functions, which are not very sensitive to the variations of $I_S$. In the case of the parallel combination, one adds two exponential functions, which are highly sensitive to the value of $V_T$.

These theoretical developments are easily applicable when the students have sorted a set of diodes. The full determination of one diode's parameters, plotting its characteristics in order to identify $I_S$ and $V_T$ takes about one our for $1^{st}$ year undergraduate students [6]. So, obtaining Figure 1 is a group work, which helps students to realize that the simple component they have under study does not have unique characteristics.

They can then choose some diodes with predetermined characteristics and see the results with a very simple electronic assembly. A possible prolongation of this work is the study of the effect of associating a small resistance to each diode (to balance the currents) in the parallel combination. An alternative is to use Light Emitting Diodes: as the luminous intensity is a function of the current, a visual inspection shows the imbalance between the currents for the parallel combination of two or more diodes (this has for example implications when designing a LED brake light).

The formulae presented in this work can easily be computed to plot the given graphs with a simple graphical calculator. This is at the level of first or second year undergraduating students.

At a higher level, one can also ask the students to establish these formulae, for example using Mathematica. These two simple examples based on very common components can help the students to appreciate the notion of validity of approximate expressions: it may help them notice that for components having non-linear behaviour, the results are strongly dependent on the type of combination.